# The SVD Beamformer: Physical Principles and Application to Ultrafast Adaptive Ultrasound

Hanna Bendjador[1], Thomas Deffieux[1], Mickaël Tanter[1]

[1] Physics for Medicine Paris, Inserm, CNRS, ESPCI Paris, Paris Sciences et Lettres University, 17 rue Moreau, 75012 Paris, France. (e-mail: hanna.bendjador@espci.fr)

*Abstract*— A shift of paradigm is currently underway in biomedical ultrasound thanks to plane and diverging waves for ultrafast imaging. One remaining challenge consists in the correction of phase and amplitude aberrations induced during propagation through complex layers. Unlike conventional line-per-line imaging, ultrafast ultrasound provides for each transmission, backscattering information from the whole imaged area. Here, we take benefit from this feature and propose an efficient approach to perform fast aberration correction based on the Singular Value Decomposition of an ultrafast compound matrix built from backscattered data for several plane wave transmissions. First, we explain the physical signification of SVD and associated singular vectors within the ultrafast matrix formalism. We theoretically demonstrate that the spatial and angular variables separation, rendered by SVD on ultrafast data, provides an elegant and straightforward way to optimize angular coherence of backscattered data. In heterogeneous media with an aberrating phase screen approximation, we demonstrate that the first spatial and angular singular vectors retrieve on one side the non-aberrated image, and on the other, the phase and amplitude of the aberration law. *In vitro* results prove the efficiency of the image correction, but also the accuracy of the aberrator determination. Based on spatial and angular coherence, we introduce a complete methodology for adaptive beamforming of ultrafast data, performed on successive isoplanatism patches undergoing SVD beamforming. The simplicity of this method paves the way to real-time adaptive ultrafast ultrasound imaging and provides a theoretical framework for future quantitative ultrasound applications.

I. INTRODUCTION

As a major medical imaging tool, ultrasound imaging constantly explores techniques to improve the image quality and extend its field of applications. The image formation obtained from ultrasonic backscattered signals, the so-called beamforming process, is actually its primary issue. It can be expressed as an inverse problem between the received ultrasonic echoes and the acoustic impedance distribution in the medium, which is easily solved when assuming the local speed of sound to be constant in the medium.

This hypothesis is accepted in all conventional clinical devices and corresponds to the conventional Delay-And-Sum beamforming, which consists in computing and correcting the time-of-flight for travel paths between each pixel and each transducer array element. In order to form high quality images at higher frame rates, ultrafast plane wave compound imaging[1] was introduced and gave rise to a wide range of applications such as Shear Wave Elastography, Ultrafast Doppler, Functional Neuroimaging or Quantitative Ultrasound imaging. However, the image quality happens to be very affected by inhomogeneities of the medium sound speed leading sometimes to non-negligible phase aberrations on propagating wavefronts [2]. Such aberrations affect the image itself but also the ensuing quantitative estimations in post-processing steps. In the last thirty years, many different approaches have been developed to address this issue but this topic remains highly relevant today despite longstanding research efforts. Measurement of the transmit echo phase [3] in Computed Ultrasound Tomography [4] informs on the arrival time of the transmitted wave front and allows correction of small aberrations compared to the PSF width. In conventional ultrasound, based on the use of backscattered signals, the ultimate goal of aberration correction is actually to recreate an equivalent point-like scatterer for each pixel of the image and retrieve the phase and amplitude distortion of its spherical backscattered echo, the so called Green's function [5]. When no bright reflector is available but rather a random distribution of Rayleigh scatterers, time-reversal of this speckle noise [6]·[7] can be used to virtually recreate an artificial ultrasonic star, whose echo retrieves easily the phase aberration laws. Though, it remains an iterative and quite complex process. Other approaches studying the spatial coherence of backscattered signals can be exploited to optimize the summation of the different plane waves in beamforming [8]. In speckle noise environment which is the vast majority of configurations assessed in Biomedical Ultrasound, Van Cittert Zernike theorem states indeed that this spatial coherence measures the focusing quality [9], which is essential for an accurate ultrasound image. Founded on the use of this theorem, coherence-based imaging methods [10] were developed to improve image quality, especially Contrast-to-Noise and Signal-to-Noise ratio, due to speckle reduction techniques. All these methods that improve spatial coherence were shown to decrease aberration across an aperture. In particular, Dahl and Trahey [11] proposed an imaging technique based on the pixel mapping of this spatial coherence of ultrasound signals. This short-lag spatial coherence method (SLSC) consists in estimating the spatial coherence between closely-spaced elements to create images demonstrating superior SNR and CNR compared to conventional ultrasound images. SLSC permits to improve the spatial coherence of backscattered signals by using only data demonstrating a strong angular coherence for different compounded transmissions without trying to estimate the aberration law [12]. This efficient approach requires the estimation of spatial coherence functions on a pixel per pixel basis whose computational cost currently hampers a real-time implementation [13], [14][15]. It was also recently expanded to angular coherence in the context of compounded plane waves.

Here, we propose a different approach to improve the ultrasonic image quality having the common aim to increase the angular coherence of signals coming from each pixel for different transmits during plane wave compound imaging. This new approach offers a fast and efficient correction method, both retrieving an optimized ultrasonic image and the estimation of the amplitude and phase of the aberrations. Importantly, we introduce the Ultrafast Compound Matrix containing images, both beamformed in the receive mode and phase delayed for transmit travel path compensation, and theoretically explain the physical meaning of its Singular Value Decomposition. In complement to former works applying SVD to subaperture or synthetic transmit data [15],[16], we hereby demonstrate a theoretical link between the SVD of beamformed data from different transmissions and the local ultrasonic aberrations. Finally, we propose a complete real-time adaptive beamforming technique for ultrafast imaging based on the application of SVD beamforming on subsets of the Ultrafast Compound Matrix corresponding to isoplanatic patches. This SVD beamformer merges Phase Aberration Correction (PAC) techniques and coherence-based imaging approaches in a unique matrix formalism implementation. Results of this novel and fast method in simulations and *in vitro* phantoms experiments are presented.

II. THEORY AND PHYSICAL INTERPRETATION OF ULTRAFAST ULTRASOUND BEAMFORMING

*A. Basic Principles of Ultrasound Imaging inverse problem*

Considering a transducer array of $N_e$ elements, and a medium containing a random distribution of *K* scatterers (Rayleigh diffusion with $K \gg N_e$), the received signal on an element *j* can be expressed as:

$$s_j(t) = \sum_{i=1}^{N_e} \sum_{k=1}^{K} \beta_k * h_{j,k}(t) * h_{k,i}(t) * e_i(t) \quad (1)$$

As the resolution of the ultrasound image is limited by the ultrasonic wavelength, it is convenient for imaging to introduce a discrete representation of the ultrasound-formed image, with a spatial pitch of the order of the ultrasonic wavelength. Thus, we define a matrix $M = \{m_{ij}\}_{i \in [1,N_x], j \in [1,N_z]}$ corresponding to the imaged medium at each pixel location *(i,j)* on a cartesian grid [$N_x$,$N_z$]. Each pixel $p = (i_0, j_0)$ contains a subset of $\{K_{i_0 j_0}\}$ scatterers. For each subset, we notice that for a scatterer *k* the travel time can be written as: $\tau_{ik} = \tau_{i_0 j_0} + \Delta \tau_k$, with $\Delta \tau_k \ll \tau_{i_0 j_0}$. Equation (1) becomes:

$$s_j(t) = \sum_{i_0=1}^{Nx} \sum_{j_0}^{Nz} \sum_{k \in \{K_{i_0 j_0}\}} \beta_{ik} * h_{j,i_0 j_0}(t - \tau_{i_0 j_0}) * \delta(t - \Delta\tau_k) * h_{i_0 j_0,i}(t - \tau_{i_0 j_0}) * \delta(t - \Delta\tau_k) * e_i(t)$$

$$= \sum_{i_0=1}^{Nx} \sum_{j_0}^{Nz} \underbrace{h_{jp}(t) * h_{pi}(t) \sum_{k \in \{K_p\}} \beta_{ik} \delta(t - 2\Delta\tau_k)}_{\gamma_{ip}(t)},$$

Thus, the received signal becomes:

$$s_j(t) = \sum_{i=1}^{N_e} \sum_{p=1}^{N_x N_z} \gamma_{ip}(t) * h_{jp}(t) * h_{pi}(t) * e_i(t) \quad (2)$$

The difference between the equation (2) and the equation (1) lies in the introduction of a scattering function $\gamma_{ip}(t)$ for each pixel, instead of a simple scalar value $\beta_{ik}$ for each scatterer. This function - characterizing the spatial dependence of the scattering in a given pixel of the image - enables the definition of a much smaller amount of Green's function $h_{ip}(t)$ as the number of pixels $N = N_x N_z \ll K$ (number of scatterers in the medium).

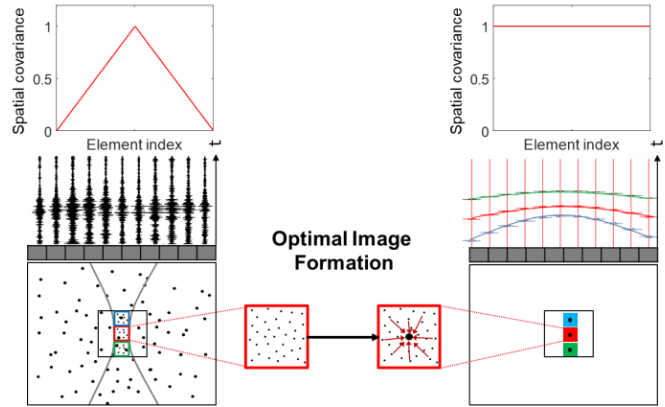

Figure 1: The ultimate goal of ultrasound imaging: speckle originates from scattering in the media. The inverse problem, or beamforming, consists in retrieving from echoes of the random distribution of scatterers inside the pixel the signature of a single point-like equivalent reflector giving access to the local Green's function of the pixel.

Depending on the nature of scatterers in each pixel of the ultrasound image, the scattering functions $\gamma_{ip}(t)$ can be very different. If pixel $p$ contains only a strong point-like reflector, the scattering function $\gamma_{ip}(t)$ is independent of the transmit element i. In that case, $\gamma_{ip}(t) = \gamma_p(t)$ and the spatial coherence function of backscattered signals coming from pixel $p$ is a square function. If the pixel $p$ contains a random distribution of Rayleigh scatterers, the function $\gamma_{ip}(t)$ is different for each transmission coming from different transmit elements and the spatial coherence function of backscattered signals coming from pixel $p$ is a triangle function according to the Van Cittert Zernike theorem [9].

The ultimate objective of ultrasound imaging is to solve the inverse problem of ultrasound propagation in order to retrieve all $\gamma_{ip}(t)$ functions for each pixel $p$. Solving this problem requires to retrieve all Green's functions $h_{ip}(t)$ relating each pixel $p$ to each element $i$ of the ultrasonic array. When pixel $p$ contains a point-like reflector, the Green's function $h_{ip}(t)$ can be retrieved by applying the concept of iterative time reversal focusing [17] or using the decomposition of the time reversal operator [18]. When pixel $p$ contains a random distribution of Rayleigh scatterers, it becomes much more complex to estimate $h_{ip}(t)$ due to the inherent speckle noise nature of backscattered signals.

It leads to a fundamental difficulty when one tries to solve the inverse problem of ultrasound imaging in speckle noise environment corresponding to the vast majority of biomedical ultrasound configurations.

Indeed, for speckle noise environments, retrieving $h_{ip}(t)$ requires to recreate a single virtual point-like reflector in each pixel $p$ of the image and separate the contribution of each pixel in the backscattered signals. Recreating such a virtual point-like reflector from speckle noise consists in increasing the spatial coherence of its backscattered signals from a triangle function to a square function. Thus, the problem of creating virtual point-like scatterers can be seen as maximizing the spatial coherence of backscattered signals coming from each focal spot. For this reason, the ultimate goal of ultrasound imaging can be seen as maximizing the spatial coherence function of backscattered signals coming from all pixels or focal spots thus enabling to retriever each Green's function associated to each pixel (

Figure 1).

*B. Fourier Domain: Beamforming and Adjoint Operator*

In the Fourier domain, convolution turns out to be a matrix product and equation (2) can be rewritten as:

$$S(\omega) = H(\omega)\Theta(\omega) \, ^tH(\omega)E(\omega) \qquad (3)$$

For a sake of clarity, we will consider in the following demonstration that $S = S(\omega_0)$, where $\omega_0$ is the central frequency of the transmitted pulse. We introduce the same notation simplification for all vectors and matrices in the following sections of the paper. Let us introduce a model of the propagation between each element of the ultrasonic array and each pixel of the image. This propagation model can be defined as a propagation operator $H_0$ as similar as possible to the true propagation operator H. The beamforming operation in the receive mode, consists in estimating

$$I = \, ^tH_0^* H \, \Theta \, ^tHE \qquad (4)$$

I is a $[N_x N_z, 1]$-vector, corresponding to the image for a transmission vector E. In the case where $H_0$ is equal to $H$, $^tH_0^* H \approx Id$ where $Id$ is the Identity Matrix and $I = \Theta \, ^tHE$.

In that case, it corresponds to the exact image of $\Theta$ for which each pixel has been multiplied by an amplitude and phase linked to the travel time path differences contained in the vector $^tHE$.

In experimental configurations due to diffraction limits, it is known that $^tH_0^*.H$ is not perfectly equal to the identity matrix, but rather estimates in each of its i[th] column the point spread function of the focusing in the i[th] pixel [19][20][21][22].

So, if $H_0$ is almost equal to $H$, we can introduce the Point Spread function Matrix Ps :

$$^tH_0^* H = \begin{bmatrix} \diagdown & & 0 \\ & \diagdown & \\ 0 & & \diagdown \end{bmatrix} = Ps \qquad (5)$$

Where all columns of $Ps$ describe the point spread functions for each individual pixel focal spot of the experimental configuration: the i[th] column vector of $Ps$ corresponds to the i[th] pixel's PSF.

*C. Conventional vs plane wave imaging*

In conventional imaging, the focusing in the transmit mode is performed by successive transmissions of transmit vectors $E_i = \, ^tH_{0,p}^*$, where $H_{0,p}$ is the $p^{th}$ column of $H_0$. For each transmit, the final image $I_m$ is the diagonal of the matrix I defined in Figure 2:

$$I_m = diag(I) = diag(\, ^tH_0^* H \, \Theta \, ^tH \, ^tH_0^*) \qquad (6)$$

In compounded plane wave imaging, the focusing is performed by a set of $N_\theta \geq 1$ plane wave transmissions. This requires a Matrix representing the change of referential basis P between the canonical space (array elements) and the plane wave transmits, so P is a $[N, N_\theta]$ matrix. The final image is thus quite similar to the conventional image described in equation (6) with a potential filtering described by $P\ ^tP^*$ when the number of plane waves is lower than the number of elements:

$$I_m = diag(I_{PW}) = diag(\ ^tH_0^*\ H\ \Theta\ ^tHP\ ^tP^*\ ^tH_0^*) \quad (7)$$

## D. The Ultrafast Compound Matrix R

We now choose to define an important matrix in compound plane wave imaging: the ultrafast compound data matrix R containing the $N_\theta$ images beamformed in the receive mode and individually time delayed pixel per pixel regarding each own transmit angle. R is thus a $[N_xN_z, N_\theta]$ matrix. It corresponds to the matrix of compounded plane wave data right before the final summation on angles, which is the last operation in ultrafast data beamforming to retrieve the coherent compounded image. We note the number of pixels $N = N_xN_z$. R can be written as:

$$R = (\ ^tH_0^*\ H\ \Theta\ ^tHP) \circ (\ ^tH_0^*P^*) \quad (8)$$

where ∘ stands for the matrix Hadamard product (
Figure 2).

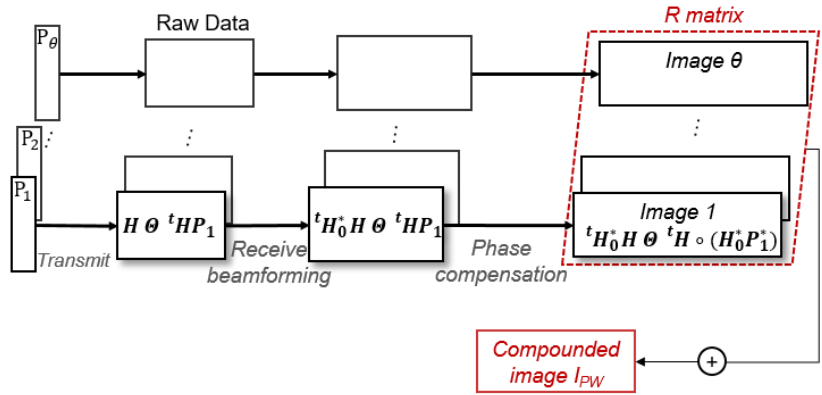

Figure 2: Matrix description of beamforming and definition of the Ultrafast Compound Matrix *R* and the Plane wave imaging matrix $I_{PW}$, and expressed as a Hadamard product.

Let us consider a homogeneous medium with an amplitude and a phase aberrating screen in the plane wave basis. The aberrator is thus a $[N_\theta, N_\theta]$ diagonal matrix A, which elements are $a = diag(A) = (a_i)_{i=1..N_\theta}$. Assuming that the aberrator does not dramatically affects transmit focusing, we can write:

$$R = (\ ^tH_0^*\ H\ \Theta\ ^tH_0\ PA) \circ (\ ^tH_0^*P^*) \quad (9)$$

The Hadamard product is a distributive operator, which is also commutative for diagonal matrices (commutative ring of $\mathcal{M}(\mathbb{C})$). Indeed, if A and B are *m\*n* matrices, and D and E are diagonal matrices of respective size *m* and *n*, it can be demonstrated that:

$D(A \circ B)E = (DAE) \circ B = (DA) \circ (BE) = A \circ (DBE)$

Thus, Equation (8) can be straightforwardly re-written as:

$$R = M . A,\ \text{where}\ M = (\ ^tH_0^*\ H\ \Theta\ ^tH_0\ P) \circ (\ ^tH_0^*P^*) \quad (10)$$

The columns of matrix *M* correspond to the ideal image of the medium, without any aberration, for each plane wave transmit. Note that the travel path differences in the transmit phase have been compensated in *M*. The ability to describe matrix R as a product between ideal images *M* and a diagonal Matrix *A* will be important in the next sections.

### E. The Local Angular Coherence Matrix $C_\vartheta$

In order to optimally correct ultrasound images, we intend to maximize the coherence of backscattered echoes between plane wave transmits, as in [23]. For this reason, we introduce a Local Angular Coherence Matrix $C_\theta$. First, we choose a control region of interest (ROI) in the image where we would like to control the angular coherence of backscattered signals. We extract the local Ultrafast Compound matrix $\tilde{R}$ of this ROI. The angular coherence of this region of interest is retrieved by calculating the covariance matrix:

$$C_\theta = {}^t\widetilde{R^*}.\tilde{R} \tag{11}$$

Figure 3 shows the computation of $C_\theta$ using experimental data in the case of a speckle environment in the region of interest. We will later discuss the size of this region of interest. We notice that the anti-diagonal terms exhibit the triangle-shaped coherence expected with Van Cittert Zernike theorem. When the same medium is now imaged through an aberrating lens, the coherence vanishes, and no submatrices show up. This witnesses the strength of the speckle coherence criterion in aberration correction. This correlation matrix informs on the angular coherence between ultrasonic images acquired with different plane wave transmits summed over all pixels of the region of interest. In speckle, as seen before, we expect it to decrease linearly from as the angle difference increases.

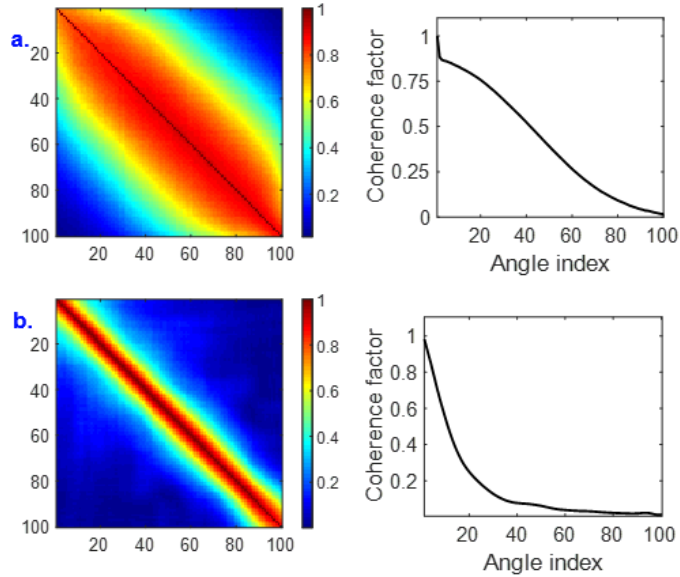

Figure 3: Measurement of the angular coherence matrix $C_\theta$ in a control region of interest using the Ultrafast compound Matrix $\tilde{R}$ extracted from R. $\tilde{R}$ is acquired in (a) non-aberrated and (b) aberrated speckle noise using 100 transmit compound angles. On the right side, diagonal terms are averaged and plotted, representing the angular coherence factor.

Interestingly, if we compute the Singular Value Decomposition of $\tilde{R}$, and re-calculate $C_\theta$ considering only the first singular vector $V_1$, it gives a uniform matrix equal to 1 (Figure 4). This means that all pixels of the singular image $V_1$ are seen the same way for all plane wave illuminations angles due to the separation of variables achieved by the SVD. The vector $U_1$ contains the amplitude and phase delays between transmits, that are required to achieve a constant coherence on the receive signal. In other words, it contains the aberration correction law of the medium acting the same way on each pixel of the region of interest. For this reason, an important point is to understand the optimal size of the control region for SVD as it is closely linked to the concept of isoplanatic patches [19][24].

### F. The Singular Value Decomposition of the Ultrafast Compound Matrix

The estimation of the angular aberration matrix $A$ is essential to perform proper adaptive imaging. Correcting the final image consists in finding the angular aberration correction vector $X$ that maximizes the angular covariance of the image $RX$. We are trying therefore to maximize the Rayleigh quotient:

$$J(X) = \frac{{}^t(RX)^* . (RX)}{X^* . X} = \frac{{}^tX^* ({}^tR^*R) X}{\|X\|}$$

One notable property of the Rayleigh quotient is that it is maximized by the first eigen vector of the matrix $({}^tR^*R)$. $({}^tR^*R)$ being a Hermitian matrix, we also know that its eigen vectors are the singular vectors of the matrix $R$. So, this demonstrates that the first singular vector of $R$ maximizes the angular covariance of the beamformed data, and thus corresponds to the aberration correction. It shows also the importance of the Singular Value Decomposition of the Ultrafast Compound Matrix.

If we approximate ${}^tH_0^*H \approx Id$, which means we consider the PSF being limited to the focal point and not affecting neighboring pixels, the previous matrix becomes:

$$M = (\Theta\ {}^tH_0 P) \circ ({}^tH_0^* P^*) = \Theta . {}^tH_0 P \circ {}^tH_0^* P^*$$

$$= \Theta . \begin{bmatrix} 1 & - & 1 \\ | & | & | \\ 1 & - & 1 \end{bmatrix} = \begin{bmatrix} \theta_1 & - & \theta_1 \\ \theta_2 & \vdots & \theta_2 \\ \vdots & \vdots & \vdots \\ \theta_N & - & \theta_N \end{bmatrix} \quad (12)$$

All columns of M are identical, and we can write:

$$R = m\ {}^ta = \begin{bmatrix} m_1 \\ \vdots \\ m_N \end{bmatrix} . [a_1 \quad \cdots \quad a_{N_\theta}] \quad (13)$$

Such angular and spatial variables separation in equation exactly corresponds to the definition of the SVD. Indeed, let's introduce the SVD of our ultrafast compound matrix R: $R = U S\ {}^tV^* = \sum_{i=1}^{N} s_{ii} u_i\ {}^tv_i^*$ where S is a diagonal matrix. As $R = m\ {}^ta$, we have here:

$$s_{11} = |\bar{m}|.|a|, \quad s_{ii} = 0\ \forall i > 1 \quad (14)$$
$$v_1 = \frac{a^*}{|a|} \quad ; \quad u_1 = \frac{\bar{m}}{|\bar{m}|}$$

$v_1$ is the phase conjugate of the aberration vector $a$. $u_1$ is the normalized image in a non-aberrated medium: all pixels are seen the same way for each plane wave transmission. In this case, it is trivial that the decomposition of $R$ is the SVD since all further singular vectors are zero, and thus orthogonal to the first one.

If we no longer consider the PSF being a spatial Dirac, and add speckle noise in the media, ${}^tH_0^*H \neq Id$ and the columns of M are no longer identical. There is thus an angular dependence of the imaged medium with respect to the insonification angle. We can introduce the averaged PSF: $\bar{m_i} = \frac{1}{N_\theta} \sum_{j=1}^{N_\theta} m_{ij}$. The elements of M can be written: $m_{ij} = \bar{m}(i) + \Delta m_i(\theta_j), \forall i = 1..N, \forall j = 1..N_\theta$.

If we consider the angular dependence of the point spread function to be reasonable, we have $\frac{|\Delta m_i|}{|\bar{m}|} \ll 1$. We can thus write:

$$R = M . A = [\bar{m}(i) + \Delta m_i(\theta_j)]_{ij} . A$$
$$= [a_j \bar{m}(i) + a_j \Delta m_i(\theta_j)]_{ij}$$
$$= \bar{m} . {}^ta + \Delta M . A \quad (15)$$

And we finally get:

$$R = |\bar{m}|.|a|.\left[\frac{\bar{m}}{|\bar{m}|}.\frac{a}{|a|} + \frac{\Delta M.A}{|\bar{m}|.|a|}\right] \quad (16)$$

with $\left|\frac{\bar{m}}{|\bar{m}|}.\frac{a}{|a|}\right| = 1$ and $\left|\frac{\Delta M.A}{|\bar{m}|.|a|}\right| \ll 1$.

Again, we know that the first vector of this SVD, with the highest eigen value, represents the best linear fit for the rows of $R$, each row being the $N_\theta$ scattering complex amplitudes of one pixel seen by each individual transmit. This first singular vector maximizes the resemblance between the rows of $R$ and the fit; in other words, the coherence between the images seen from different angles. Physically, this is equivalent to a maximization of the energy of $R$ rows which corresponds to the highest energy of all singular vectors. As we know, due to the speckle noise features that the term $\Delta M.A$ is negligible compared to $\bar{m}.\,^t a$, we can again identify:

$$s_{11} = |\bar{m}|.|a| \quad ; \quad u_1 = \frac{\bar{m}}{|\bar{m}|} \quad ; \quad v_1 = \frac{a^*}{|a|}.$$

The first singular vector of $R$ is containing in $v_1$ the phase conjugate of the aberrator. $u_1$ contains the normalized image of the medium where the aberration has been corrected, with all pixels behaving as individual point-like reflectors with no angular dependence in transmission, as the uniform coherence matrices witness in Figure 4 below.

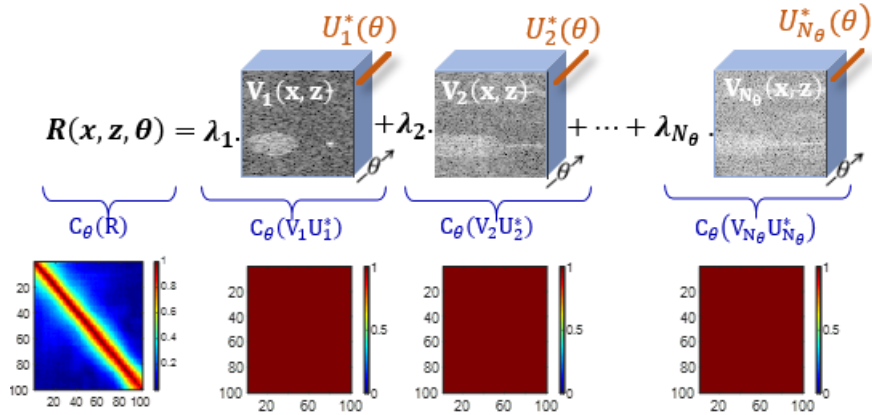

Figure 4: Singular Value Decomposition of the Ultrafast Compound Matrix. Angular Coherence matrix of R shows speckle decorrelation, whereas singular vectors exhibit individually an optimized angular coherence.

Strikingly, the Singular Value Decomposition of the Ultrafast Compound matrix gives access, with its first singular vector, to the knowledge of the aberrating screen. We notice that selecting the first vector out of the SVD is equivalent to re-phasing the images with the phase conjugate of the aberrator screen in order to create an angular coherence for each point of an aberrated medium. This is the definition of creating a virtual reflector in each pixel location. This is actually the mathematical explanation for the bright reflector virtual creation in speckle noise described in [6]: backscattered signals are re-phased and summed "by hand" to exhibit a virtual point like reflector out of speckle noise. The only difference in these approaches lies in the nature of signals, here beamformed data are used for rephasing, whereas in the other case, the point source generation was made from the RF echoes. The SVD beamforming proposed here is a straightforward and non iterative solution for this complex problem.

One should note that the aberration was assumed to have an effect only in the transmit mode, for a sake of simplicity in the former mathematical developments. The aberration in the receive mode can be taken into account by re-writing Equation (8) in:

$$R = \left(\,^t H_0^* P_0^* \Theta A\,^t P_0\,^t HPA\right) \cdot \left(\,^t H_0^* P^*\right)$$

Where $P_0$ is the orthonormal $[N_e, N_e]$-matrix, changing from canonical to plane wave basis with $rank(P_0) = N_e$. In ultrafast imaging, $rank(P_0)$ is often much greater than $rank(P)$ as $N_e \gg N_\theta$ in most cases. Then, likewise, we can write :

$R = M.A = \bar{m}.\,^t a + \Delta M.A$, and the SVD retrieves as well the aberration matrix.

## III. METHODS

### A. Acquisitions

Ultrasound acquisitions were performed using a 256-channel programmable research scanner (Verasonics Research Systems). We developed customized sequences to drive a 192-element linear probe, at a central frequency of 6.25 MHz and pitch size of 0.2 mm (SL10-2 probe, Supersonic Imagine, Aix-en-Provence, France). Plane waves steered at angles between -18° and +18° were transmitted to insonify the media. In order to guarantee the highest level of information, 100 angles at Pulse Repetition Frequency (PRF) = 10 kHz were acquired along with subsets of steering angles for higher frame rate images. A phantom containing reflecting pins, hyperechoic and anechoic cysts (CIRS 054GS – 1540 m/s) was used to assess our method, successively without and with a shaped-surface silicon aberrating lens.

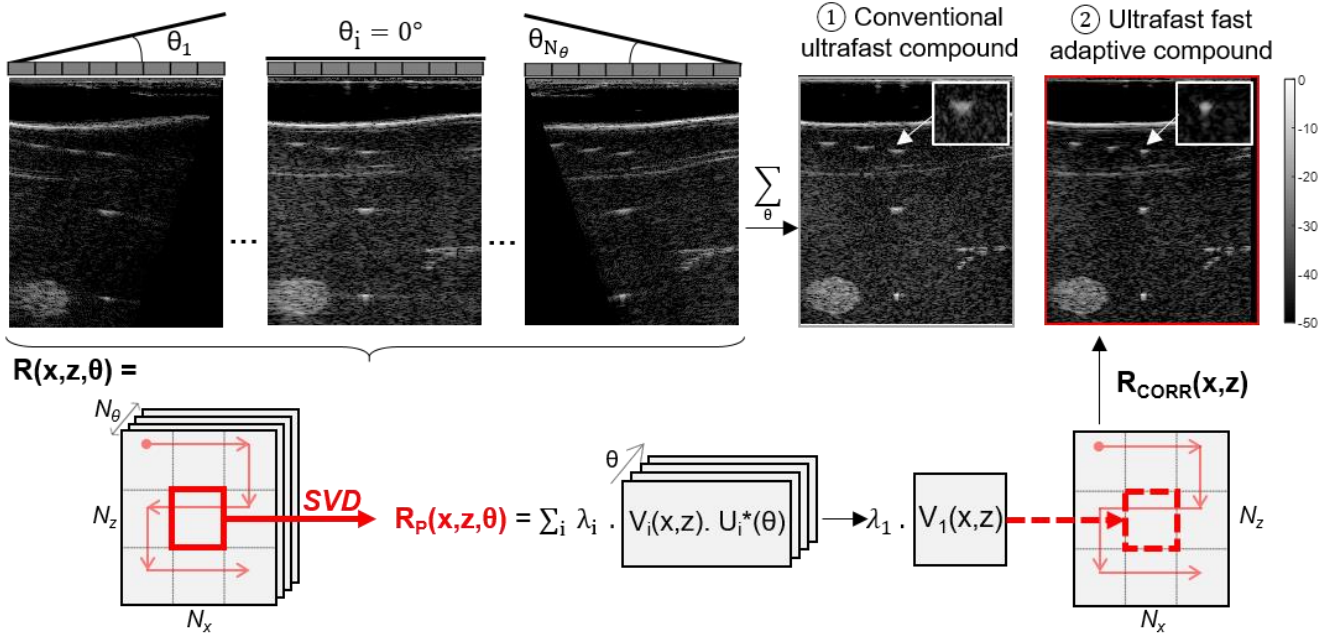

Figure 5: Fast aberration correction and filtering with SVD. Ultrafast compound matrix R retrieves IQ data beamformed for each transmit angle. Isoplanatic patches are defined within the imaged area. SVD is performed on each patch of the image, and the first eigen vector gives directly the rephased image. Final corrected image is computed by assembling corrected patches.

### B. Correction method

Classical beamforming - at $dx = \lambda$, $dz = \lambda/2$ - was performed on plane wave data for each steered transmit. Resulting IQ signals were stored in a 3D-matrix R: the *ultrafast compound matrix* containing (x,z)- images for each angle $\theta$ (Figure 5①). Summation along the third dimension, gives simply the conventional plane wave compounded B-mode image.

Matrix R was reshaped into a 2D Casorati Matrix form with a $[N_x \, N_z, \, N_\theta]$ dimension and the Singular Value Decomposition was performed on delimited spatial ROIs called "control patches", each corresponding to a smaller Ultrafast Compound matrix $\tilde{R}$. We can adapt the dimension of the control ROIs with respect to the isoplanatism limited extension induced by the aberrator. SVD of each $\tilde{R}$ matrix is performed and provides the separation of spatial and angular variables. For each singular value, the SVD consists actually in the product of an image $V(x,z)$ and an angular vector $U^*(\theta)$ with eigen values $\lambda$. The decomposition is filtered, keeping only the image with the highest singular value. For each angle, the corrected image of each ROI is thus corresponding to $\lambda_1 . V_1(x,z)$. Summing data on the different angles retrieves this corrected image (Figure 5②). It has been rephased with the complex $U_1(\theta)$ containing the phase and amplitude corrections. As described in part II., the first eigen vector guarantees a maximization of the angular coherence. It is constant square function equal to 1, since all angles "see" the same outcome from the object. We studied three experimental cases for comparison: the classical ultrafast compound imaging in homogeneous medium, the SVD beamforming using experimental data incorporating a *numerical* aberration and finally experimental data containing a *physical* aberration. The classical ultrafast compound in homogeneous medium was considered as the goal to reach for the aberrated images correction by the SVD beamformer.

We call *numerical* aberration, the introduction at emission on acquired data, of a known angular delay law (Figure 6) and we evaluated the consistency between the phase aberration determined with our method, and the expected one. We also introduced *physical* aberrations: between the probe and the phantom, we placed a silicon aberrating lens and applied the SVD beamformer to compensate the defaults in the image (Figure 7).

Quantification of the image quality can be found in the angular coherence matrix, which mathematically becomes constant and maximal between all angles after correction. Though, we can also provide classical estimators such as lateral resolution – computed by the half width lateral resolution of a reflecting pin intensity, and contrast defined by: $= \frac{\mu_i}{\mu_o}$, where $\mu_i = \overline{s_i}^2$ and $\mu_o = \overline{s_o}^2$ are the mean square intensities respectively inside and outside an anechoic cyst.

IV. RESULTS

The angular coherence in speckle for a casual B-mode image, calculated for a control region, is proven to be a block matrix, each block being the angular coherence between a given control patch and itself (diagonal) or with another control patch. Diagonal blocks, corresponding to angular coherence of $\tilde{R}$ matrix exhibit interestingly a triangle profile: covariance is maximal along the diagonal – auto-coherence - and vanishes linearly down to zero when the angular spacing increases (Figure 3). Van Cittert Zernike theorem stipulates actually that the shape of the spatial coherence is driven by the inverse Fourier transform of the focal spot intensity (itself proportional to sinc² = FT(triangle) ). Indeed in speckle, one obtains a linear decrease in coherence as much as the lag between elements increases. Also, scattering creates a decorrelation between signals of two different points that would justify the weak angular coherence between isoplanatic patches. When it comes to an aberrated image, speckle is highly decorrelated and backscattered signals angular coherence sinks very quickly from an angle to the next one. After SVD correction and filtering of the image, this effect is totally recovered, and even further since each block is constant equal to 1. Even if this result seems certain mathematically, it illustrates that the SVD provides an image, for which all pixels are being seen the same way by all transmit angles.

Compared to the raw image, the numerically aberrated image showed a strong degradation (Figure 6) that is fully recovered after SVD correction. The lateral resolution is effectively improved from 1.02 mm to 0.88 mm. Also, the contrast on an anechoic cyst is enhanced by 11.7 dB+/- 1.1 dB. The aberration law, extracted from the phase of the angular singular vector is shown to be very consistent with the introduced phase change (r² = 99%).

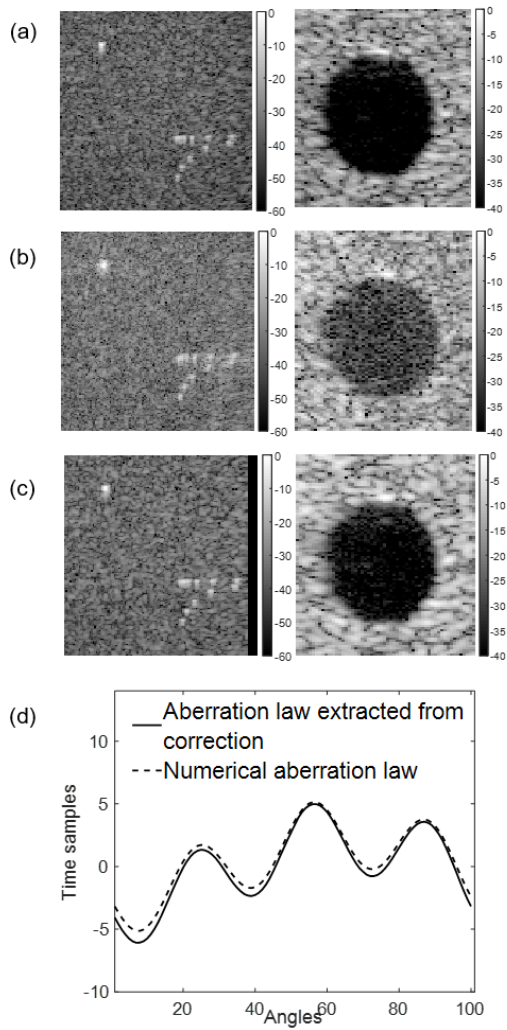
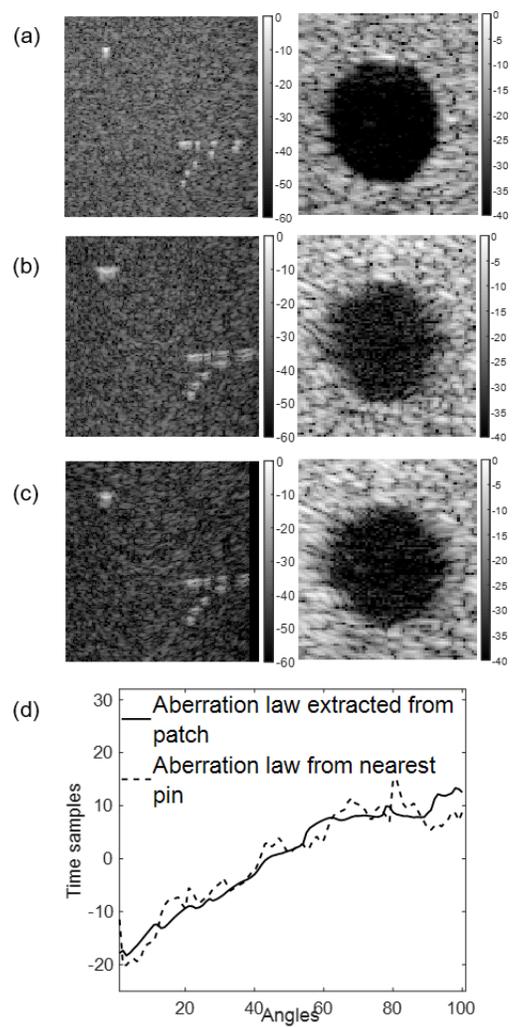

Figure 6: (a) Bmode images of a CIRS 054-GS phantom reflecting pins and anechoic cyst. (b) Images after numerical aberration at emission. (c) Images after SVD-beamformer. (d) Aberrating laws: numerical and extracted from SVD-beamforming.

Figure 7: (a) Bmode images of a CIRS 054-GS phantom reflecting pins and anechoic cyst. (b) Images after propagation through an aberrating lens. (c) Images after SVD-beamformer. (d) Aberrating laws: extracted from SVD-beamforming on a patch, and from the closest pin echo.

These improvements of the SVD beamformer are visible as well on physically aberrated data (Figure 7). The difference lies in the process of SVD decomposition that is performed on isoplanatic patches due to the thickness of the aberrator. Reflecting pins resolution improved from 2.1mm to 1.3mm, and contrast by 8.5 dB+/- 1.9 dB. We compared here, the aberration law extracted from patch correction with its true estimation obtained by using the backscattered wavefront coming from the closest pin whose aberration. We found again a very good agreement ($r^2$ = 95%) between those two angular aberration laws. In the case of a thick aberrating lens, smaller isoplanatic patches were used to perform the correction, in order to keep the aberration as uniform as possible on the patch spatial extension. An interesting issue remains in the optimization of this patch size in order to perform the most efficient possible correction. In order to investigate this point, we designed patches around a central point of five different sizes, and performed correction on each of them with SVD beamforming (Figure 8).

The phase of the first angular eigen vector was shown to be consistent for the three smaller patches. That means that for this patch sizes, the patch is seeing the same aberrator and the SVD beamformer provides an optimal correction and estimation of the aberration. When increasing the patch size, we observe a change in the phase appearance towards a hyperbolic shape. The spatial extent of patch 4 becomes too large to ensure a correct solution for the SVD beamformer. This shape translates a mistaken sound speed during beamforming. We are thus able to identify a typical patch size for aberration correction. It is actually the highest size for which the phase laws remain consistent. Figure 8 clearly illustrates that the patch size should be correctly chosen: for a too small patch, the number of pixels is not sufficient for the SVD to extract the correct aberration law. For too large patch size, the aberration can no longer be considered similar for all pixels of the patch, and it is logical that the SVD performance degrades. This optimization of the patch size can be done by looking at the singular value distribution of the SVD for different sizes. As seen in Figure 8 (c), the singular value distribution drops rapidly to zero for a correct patch size, whereas in the case of too small patches, the slope between the first and lower singular values is less steep. The ratio between the first and second singular value of the Ultrafast Compound Matrix can be considered as an interesting parameter to choose the size of isoplanatic patch. In the case of the time reversal operation $H^tH^*$, such ratio was shown of particular importance as it affects the convergence of iterative time reversal processing [18], [25].

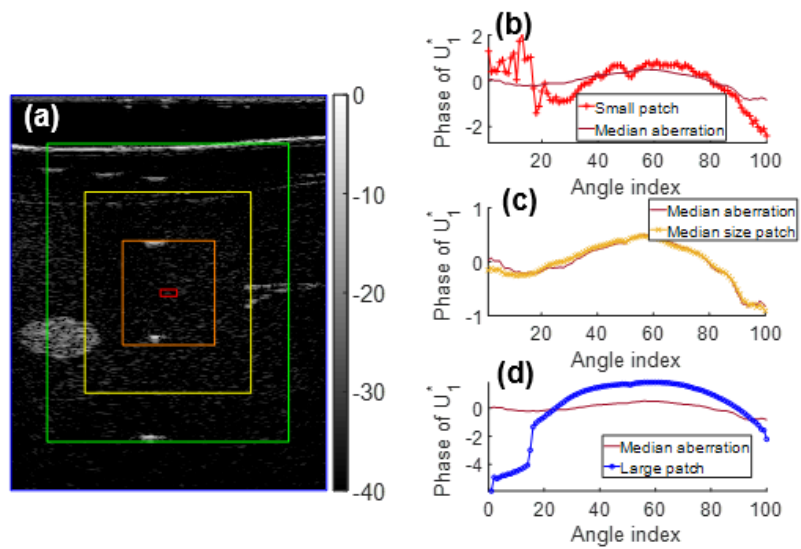

Figure 8: (a) B-mode image showing different patch sizes used for SVD beamforming [red: 5λ × 5λ, orange: 25λ × 75λ, yellow: 50λ × 150λ, green: 70λ × 220λ, blue: 100λ × 300λ]. *"Median aberration"* stands for the mean of angular aberration laws within the median range of patch sizes. (b) Angular aberration laws extracted from SVD beamformer for a small patch compared to the median aberration. (c) For a median size patch. (d) For a large patch.

## V. DISCUSSION

The results presented above demonstrate that the SVD beamformer is a reliable and efficient technique to correct images and retrieve simultaneously the phase aberration law from ultrafast data acquired using plane wave compounding. In vitro experiments in tissue mimicking phantoms have shown both a 11.7 dB+/- 1.1 dB rise in contrast for simulated aberrations and a 8.5 dB+/- 1.9 dB rise for physical aberrations. The lateral resolution was also improved in both cases respectively from 1.02mm to 0.8mm in the simulated aberration case, and outstandingly from 2.1mm to 1.3mm with the physical aberration. Another simple way to assess the accuracy of our method in retrieving the aberration pattern was to beamform raw data while correcting the phase and amplitude aberrations determined by SVD, and to re-perform the SVD beamforming process on this new set of corrected data. As expected, this resulted in a flatten phase distribution, as if no aberration had distorted the backscattered signals. This showed also that a single round of correction was required whereas existing methods often imply iterative processes. Of particular interest is the simplicity and straightforward implementation of this SVD beamforming. Using parallel computation devices, and high-end processors, the implementation time of our method is significantly low enough to perform real-time aberration correction. It depends on the number of transmit angles in the data, but also on the size of the isoplanatic patches for Singular Value Decomposition. We showed in Table 1, that for a reasonable amount of plane waves, and for a patch size of tens of λ in axial and lateral directions, the computation speed is yet high enough to correct in real-time the beamformed data which is a major advantage of this SVD beamformer.

| Processing | Time for processing (s) | | | |
|---|---|---|---|---|
| Number of patch(es) | 1 | 9 | 25 | 100 |
| Size of patch (pixels) | 592*192 | 196*63 | 117*37 | 58*18 |
| Beamforming 100 angles | 0.000036 | | | |
| SVD-correction 100 angles | 0.50 | 0.41 | 0.34 | 0.50 |
| SVD-correction 10 angles | 0.025 | 0.018 | 0.023 | 0.045 |
| SVD-correction 5 angles | 0.012 | 0.010 | 0.014 | 0.029 |

Table 1: Calculation times for an image - beamformed at λ/2 - acquired at a central frequency of 6.25 MHz and a sampling frequency of 25 MHz, at 5cm depth with a 192-element probe. Calculations were computed on a PC with an IntelCorei7-5820K CPU (3.30GHz), 32 Go RAM, on a Graphics Processing Unit: GeForce GTX 1080 TI.

With a deeper interest into parallel computing of various algorithms, the computation time could be even further reduced, allowing to consider novel applications such as real-time adaptive functional imaging, motion-correction on cardiac imaging, or transcranial Doppler imaging of the brain. For transcranial imaging, it could seem limitative that our method only corrects in emission. Though, due to time-reversal symmetry of the wave equation, the aberration is also symmetric. It is therefore possible to correct in emission, and in reception by re-beamforming the data while taking the aberration delay into account in both propagation ways.

The optimization of the number of isoplanatic patches and their overlapping for the final image reconstruction was not discussed in the manuscript but it is also an interesting problem with room of further improvements. The size of the isoplanatic region depends on geometric parameters (such as depth, aperture) and aberration characteristics (such as correlation length). To date, we decided to choose a typical 10λ × 70λ patch size for the SVD beamforming corresponding to classical aberrations and geometry in medical ultrasound, but this could be refined in further works. A 50% overlapping between patches was chosen and the final image was reconstructed by choosing for each pixel the mean value of the same pixel in the several overlapping patches. This enables to avoid spatial discontinuities at the frontier between patches. Other weighting methods before summation over the different patches could be studied in further works in order to optimize the computation time.

In our study, we considered a simple scattering regime for the backscattered signals. This hypothesis is sufficient in most configurations of biomedical ultrasound. Nevertheless, the discrimination between signals resulting from simple and multiple scattering could be performed at a non-negligible supplemental computational cost before our SVD beamforming method by studying the propagation matrix singular vectors [26].

In the case of a non-aberrated medium, for a correct image, the first angular eigen vector out of SVD exhibit obviously a constant phase at 0. Though, if there is a mismatch between the beamforming sound speed and the medium sound speed, the phase immediately shows a hyperbolic trend – either convex or concave depending on the sign of the error. So, the phase of the eigen value straightforwardly informs on the sound speed accuracy in the patch of interest. The framework of this SVD beamformer processing is thus particularly adapted for quantitative imaging, and particularly acoustic sound speed mapping.

## VI. CONCLUSION

The SVD Beamformer provides a fast and adaptive method for beamforming whilst correcting phase and amplitude aberrations. On a theoretical point of view, applying the SVD on a particularly suited ultrafast compound Matrix R gives a straightforward solution to the aberration correction problem. This technique for adaptive ultrasound imaging reunites Phase Aberration Correction and Coherence-based imaging in a single correction operation. Selecting the first image singular vector is sufficient to maximize the angular coherence between transmitted insonifications and enhances the image quality, without compromising very high frame rate imaging. We propose for the first time a physical meaning to the mathematical SVD operation. Future work will focus on the application of this technique for quantitative ultrasound imaging.

REFERENCES

[1] G. Montaldo, M. Tanter, J. Bercoff, N. Benech, and M. Fink, "Coherent Plane Wave Compounding for Very High Frame Rate Ultrasonography and Transient Elastography," *IEEE Trans. Ultrason. Ferroelectr. Freq. Control*, vol. 56, no. 3, pp. 489–506, 2009.

[2] G. E. Trahey and S. W. Smith, "Properties of acoustical speckle in the presence of phase aberration part I: first order statistics," *Ultrason. Imaging*, vol. 10, no. 1, pp. 12–28, 1988.

[3] M. Jaeger, E. Robinson, H. G. Akarçay, and M. Frenz, "Full correction for spatially distributed speed-of-sound in echo ultrasound based on measuring aberration delays via transmit beam steering," *Phys. Med. Biol.*, vol. 60, no. 11, pp. 4497–4515, 2015.

[4] M. Jaeger, G. Held, S. Peeters, S. Preisser, M. Grünig, and M. Frenz, "Computed Ultrasound Tomography in Echo Mode for Imaging Speed of Sound Using Pulse-Echo Sonography: Proof of Principle," *Ultrasound Med. Biol.*, vol. 41, no. 1, pp. 235–250, 2015.

[5] M. O'Donnell and S. W. Flax, "Phase-Aberration Correction Using Signals From Point Reflectors and Diffuse Scatterers: Measurements," *IEEE Trans. Ultrason. Ferroelectr. Freq. Control*, vol. 35, no. 6, pp. 768–774, 1988.

[6] G. Montaldo, M. Tanter, and M. Fink, "Time reversal of speckle noise," *Phys. Rev. Lett.*, vol. 106, no. 5, pp. 1–4, 2011.

[7] B. F. Osmanski, G. Montaldo, M. Tanter, and M. Fink, "Aberration correction by time reversal of moving speckle noise," *IEEE Trans. Ultrason. Ferroelectr. Freq. Control*, vol. 59, no. 7, pp. 1575–1583, 2012.

[8] N. Q. Nguyen and R. W. Prager, "A Spatial Coherence Approach to Minimum Variance Beamforming for Plane-Wave Compounding," *IEEE Trans. Ultrason. Ferroelectr. Freq. Control*, vol. 65, no. 4, pp. 522–534, 2018.

[9] R. Mallart and M. Fink, "The van Cittert–Zernike theorem in pulse echo measurements," *J. Acoust. Soc. Am.*, vol. 90, no. 5, pp. 2718–2727, 1991.

[10] M. A. Lediju, G. E. Trahey, B. C. Byram, and J. J. Dahl, "Short-lag spatial coherence of backscattered echoes: Imaging characteristics," *IEEE Trans. Ultrason. Ferroelectr. Freq. Control*, vol. 58, no. 7, pp. 1377–1388, 2011.

[11] J. Dahl, "Coherence beamforming and its applications to the difficult-to-image patient," *IEEE Int. Ultrason. Symp. IUS*, 2017.

[12] L. Nock, G. E. Trahey, and S. W. Smith, "Phase aberration correction in medical ultrasound using speckle brightness as a quality factor.," *J. Acoust. Soc. Am.*, vol. 85, no. 5, pp. 1819–33, 1989.

[13] N. Riaz, S. L. Wolden, D. Y. Gelblum, and J. Eric, "Coherent Flow Power Doppler (CFPD): Flow Detection using Spatial Coherence Beamforming," *IEEE Trans Ultrason Ferroelectr Freq Control.*, vol. 118, no. 24, pp. 6072–6078, 2016.

[14] J.-L. Gennisson *et al.*, "Robust sound speed estimation for ultrasound-based hepatic steatosis assessment," *Phys. Med. Biol.*, vol. 62, no. 9, pp. 3582–3598, 2017.

[15] A. Aubry and A. Derode, "Detection and imaging in a random medium: A matrix method to overcome multiple scattering and aberration," *J. Appl. Phys.*, vol. 106, no. 4, 2009.

[16] F. Vignon, J. S. Shin, S. W. Huang, and J. L. Robert, "Adaptive ultrasound clutter rejection through spatial eigenvector filtering," *IEEE Int. Ultrason. Symp. IUS*, no. Table 1, pp. 31–34, 2017.

[17] G. Montaldo, M. Tanter, and M. Fink, "Real time inverse filter focusing by iterative time reversal," *J. Acoust. Soc. Am.*, vol. 112, no. 5, pp. 2446–2446, 2013.

[18] C. Prada and M. Fink, "Eigenmodes of the time reversal operator: A solution to selective focusing in multiple-target media," *Wave Motion*, vol. 20, no. 2, pp. 151–163, 1994.

[19] M. Tanter, J.-L. Thomas, and M. Fink, "Focusing and steering through absorbing and aberrating layers: Application to ultrasonic propagation through the skull," *J. Acoust. Soc. Am.*, vol. 103, no. 5, pp. 2403–2410, 1998.

[20] M. Tanter, J.-L. Thomas, and M. Fink, "Time reversal and the inverse filter," *J. Acoust. Soc. Am.*, vol. 108, no. 1, pp. 223–234, 2002.

[21] M. Tanter, J.-F. Aubry, J. Gerber, J.-L. Thomas, and M. Fink, "Optimal focusing by spatio-temporal inverse filter. I. Basic principles," *J. Acoust. Soc. Am.*, vol. 110, no. 1, pp. 37–47, 2002.

[22] J.-F. Aubry, M. Tanter, J. Gerber, J.-L. Thomas, and M. Fink, "Optimal focusing by spatio-temporal inverse filter. II. Experiments. Application to focusing through absorbing and reverberating media," *J. Acoust. Soc. Am.*, vol. 110, no. 1, pp. 48–58, 2002.

[23] Y. L. Li and J. J. Dahl, "Angular coherence in ultrasound imaging: Theory and applications," *J. Acoust. Soc. Am.*, vol. 141, no. 3, pp. 1582–1594, 2017.

[24] F. Chassat, "Theoretical evaluation of the isoplanatic patch of an adaptive optics system working through the atmospheric turbulence," *J. Opt.*, vol. 20, no. 1, pp. 13–23, 1989.

[25] C. Prada, J. Thomas, and M. Fink, "The iterative time reversal process: Analysis of the convergence," *J. Acoust. Soc. Am.*, vol. 97, no. 1, pp. 62–71, 2005.

[26] A. Aubry and A. Derode, "Singular value distribution of the propagation matrix in random scattering media," *Waves in Random and Complex Media*, vol. 20, no. 3, pp. 333–363, 2010.